%% file: main.tex
\documentclass[conference,compsoc]{IEEEtran}

\usepackage[T1]{fontenc}
\usepackage[utf8]{inputenc}
\usepackage{cite}
\usepackage{amsmath,amssymb}
\usepackage{booktabs}
\usepackage{graphicx}
\usepackage{xcolor}
\usepackage{listings}
\usepackage{needspace}
\usepackage{placeins}
\usepackage[skins,breakable,listings]{tcolorbox}
\usepackage{tikz}
\usetikzlibrary{arrows.meta,positioning}
\usepackage{url}
\usepackage[hidelinks]{hyperref}

\newcommand{\sys}{Collective Integrity Contracts}
\newcommand{\shortsys}{CIC}
\newcommand{\papertitle}{Gerrymandering the Warp: Non-Control-Data Attacks on CUDA Collective Decisions}

\lstdefinestyle{cuda}{
  basicstyle=\ttfamily\footnotesize,
  columns=fullflexible,
  keepspaces=true,
  showstringspaces=false,
  breaklines=true
}
\newtcblisting{codetiny}{
  listing only,
  listing options={style=cuda},
  enhanced,
  breakable,
  colback=black!2,
  colframe=black!35,
  boxrule=0.25pt,
  arc=1pt,
  left=0.45em,
  right=0.25em,
  top=0.25em,
  bottom=0.25em,
  boxsep=0.15em,
  before skip=0.45em,
  after skip=0.45em
}
\newtcblisting{codeatom}[1][12]{
  listing only,
  listing options={style=cuda},
  enhanced,
  unbreakable,
  before={\Needspace{#1\baselineskip}},
  colback=black!2,
  colframe=black!35,
  boxrule=0.25pt,
  arc=1pt,
  left=0.45em,
  right=0.25em,
  top=0.25em,
  bottom=0.25em,
  boxsep=0.15em,
  before skip=0.45em,
  after skip=0.45em
}

\title{\papertitle}

\author{Igor Santos-Grueiro\\
International University of La Rioja}
\begin{document}

\maketitle

\begin{abstract}
\input{sections/00_abstract}

\end{abstract}

\input{sections/01_introduction}
\input{sections/02_background}
\input{sections/03_participation_integrity}
\input{sections/04_threat_model}
\input{sections/05_attacks}
\input{sections/06_evaluation}
\input{sections/07_binding_participation_authority}
\input{sections/09_discussion_limitations}
\input{sections/10_related_work}
\input{sections/11_conclusion}

\bibliographystyle{IEEEtran}
\bibliography{refs}

\appendices
\input{sections/appendix}

\end{document}

%% file: sections/00_abstract.tex
CUDA collective operations often sit on security decision paths: votes accept batches, reductions aggregate evidence, shuffles select representatives, and barriers order checked state before use. Such decisions depend not only on computed values, but also on which lanes are represented, what evidence they contribute, which lane speaks for the group, and which checked state reaches commit. We identify this participation metadata as decision-making non-control data.

We define Collective Semantic Corruption (CSC), a non-control-data attack family in which range-valid masks, predicates, source lanes, descriptors, group labels, or epochs cause a CUDA-conforming collective to authorize a decision over the wrong membership, contribution, role, or validation-to-use state. The kernel reaches the intended collective site and executes the expected primitive; the primitive represents the wrong authority set.

We model CSC with a site-local participation-authority contract. A protected collective derives, recomputes, checks, or freezes membership, contribution, role, and temporal state before authorization. We evaluate CSC across NVIDIA CUDA collective primitives, trigger channels, compact workload-style kernels, reduced idiom bridges, and admission-guard harnesses. In a CUDA-defined contract-conformance suite spanning the four authority dimensions, corrupted participation metadata causes a trusted-reference mismatch in 102/102 instances, while hardened variants preserve that reference in 102/102. We report 13 synchronization-sensitive instances separately. We then introduce Collective Integrity Contracts (CIC), a wrapper discipline that binds participation metadata before collective use. For CUDA collective decisions, security depends on both the values computed and the participants represented.

%% file: sections/01_introduction.tex
\section{Introduction}

Non-control-data attacks subvert security logic by corrupting the data on
which a program's decisions depend, while preserving the intended control
flow. In scalar software, such data can include identities, policy flags,
configuration values, authorization bits, and input objects
~\cite{chen2005noncontrol,schlesinger2014modular,hu2016dop}. CUDA applications
introduce another security-relevant object: participation metadata. A
collective decision can depend not only on the values computed by lanes, but
also on which lanes are represented, which predicates contribute, which lane
speaks for the group, and whether checked state remains bound to later use.

CUDA collectives often appear on security-relevant decision paths. Votes accept
or reject batches, ballots and reductions aggregate evidence, shuffles select
representatives, barriers order shared-state consumption, and cooperative
groups define domains for group work. At such sites, the CUDA synchronization
mask and the application-level semantic authority set are distinct objects.
The synchronization mask specifies which non-exited lanes named by the mask
must execute the primitive for CUDA-defined behavior. The semantic authority
set specifies which principals the application decision is allowed to
represent. A kernel can therefore execute the intended primitive at the
intended program point, satisfy the CUDA call discipline, and still authorize a
decision over the wrong authority set.

In this paper, we call an execution \emph{CUDA-conforming} when all non-exited lanes required by the primitive's call mask execute the expected primitive at the expected program point. The \emph{CUDA-defined core} is the evaluated subset that satisfies this condition and excludes synchronization-sensitive or
undefined-behavior-sensitive cases.

Consider a warp-level admission check in which each lane validates one record
and a warp vote accepts the batch only if all intended records are valid. If
mutable metadata excludes the failing lane from the semantic participant set,
the vote still executes at the expected site and returns acceptance over the
represented lanes. The program counter, intrinsic, synchronization mask, and
value ranges remain valid. The corrupted non-control data is the
\emph{participation authority}: the metadata that determines which lanes the
collective decision is allowed to represent.

We identify this failure as \emph{Collective Semantic Corruption} (CSC). A CSC
instance corrupts program-visible data that defines membership, contribution,
role, or validation-to-use binding for a security-relevant collective. The
corrupted object may be a range-valid mask, predicate, source lane,
descriptor, group label, rank, valid count, or epoch. The kernel reaches the
intended collective site and executes the expected CUDA primitive; the result
authorizes a decision over invalid participation authority.
CSC is a program-level failure: CUDA programs that treat mutable participation
metadata as authority for protected collective decisions expose this attack
surface. Core CSC instances preserve that CUDA call discipline. The corrupted object is application-level participation metadata consumed by the primitive: a semantic participant predicate, contribution, role, descriptor, or validation to use binding.

Participation-authority corruption can enter through several data paths.
Adversarial metadata may arrive in a host request as a syntactically valid
descriptor, count, group label, or leader field. Stale metadata may arise when
one phase validates a descriptor and a later collective reloads changed state.
A device-side memory bug may overwrite a valid flag, mask, rank, descriptor,
or epoch before the collective consumes it. These channels differ in cause,
but share the same use-site failure: mutable metadata is treated as authority
for a collective decision.

We model each security-relevant collective with a site-local
\emph{participation-authority contract}. The contract names the trusted
membership, contribution, role, and temporal state that must be derived,
checked, recomputed, or frozen before authorization. This separates
range-valid metadata from authority-valid metadata. A mask can contain legal
lane bits, a source lane can be in range, and a descriptor can be in bounds,
while still naming the wrong authority for the collective.

The NVIDIA CUDA evaluation has a main contract-conformance suite plus
additional evidence blocks. The main suite includes 102 core CUDA instances
covering membership, contribution, role, and timing. In each case,
corrupted participation metadata breaks the trusted reference, while hardened
variants keep it intact. A separate set of 13 synchronization-sensitive
instances lies outside the core claim. Fuzzers, admission-guard harnesses,
negative controls, checker triage, and disclosure follow-up offer extra
support. \sys{} (\shortsys{}) is a CUDA wrapper discipline that derives,
checks, recomputes, or freezes participation metadata before a collective uses
it.

Our paper makes the following contributions:
\begin{itemize}
    \item \textbf{Participation authority.} We identify CUDA participation
    metadata as decision-making non-control data: a collective can execute
    correctly under CUDA's call discipline while representing the wrong
    authority set.
    \item \textbf{Contract model.} We formalize a site-local
    participation-authority contract for membership, contribution, role, and
    validation-to-use binding.
    \item \textbf{CUDA evidence.} We evaluate CSC through a CUDA-defined
    contract-conformance suite, trigger-channel variants, fuzzers,
    admission-guard harnesses, reduced idiom bridges, and a disclosed
    public-API follow-up, with synchronization-sensitive instances reported
    separately.
    \item \textbf{Binding discipline.} We introduce \shortsys{}, a wrapper
    discipline that derives, checks, recomputes, or freezes participation
    metadata before collective use.
\end{itemize}


%% file: sections/02_background.tex
\section{Background: Non-Control Data and CUDA Collective Decisions}
\label{sec:background}

\subsection{Non-Control-Data Attacks}

Non-control-data attacks corrupt the data on which security decisions depend
while preserving the program's intended control flow, a boundary
complementary to control-flow and code-pointer integrity
~\cite{abadi2005cfi,kuznetsov2014cpi}. Classic targets include identities,
policy flags, configuration state, input objects, and other decision
data~\cite{chen2005noncontrol,hu2016dop}. Defenses such as
data-flow integrity and modular non-control-data protection can protect this
state once a policy identifies the objects whose integrity matters
~\cite{castro2006dfi,schlesinger2014modular}.

CUDA collective decisions introduce another policy object: participation
metadata. A security-relevant collective may depend on a mask, predicate,
source lane, group label, rank, descriptor, valid count, or epoch. These values
determine which lanes a decision represents, what those lanes contribute, which
lane may speak for the group, and whether previously checked state remains
bound to later use.

\subsection{CUDA Collectives as Decision Procedures}

CUDA programs organize execution into grids, thread blocks, and warps. A warp
contains lanes that often progress through a common instruction stream, but
the instruction stream is only one component of execution state. A lane may be
active, predicated off, exited, masked out, or outside the domain selected by a
collective operation. Consequently, the application-level decision represented
by a collective can depend on program-selected domains, predicates, roles, and
contribution sets~\cite{nvidiaCudaGuide,nvidiaPtxIsa,nvidiaBestPractices,
habermaier2012simt}.

Warp-level CUDA primitives make participation explicit. Operations such as
\texttt{\_\_all\_sync}, \texttt{\_\_any\_sync},
\texttt{\_\_ballot\_sync}, \texttt{\_\_shfl\_sync}, and
\texttt{\_\_syncwarp} execute over a mask or domain chosen by the program
~\cite{nvidiaCudaGuide,nvidiaPtxIsa,linCudaWarpPrimitives}. These operations
often sit directly on application decisions. A vote may accept a batch only if
all intended records validate. A reduction may admit a candidate only if valid
contributions reach a threshold. A shuffle may select the lane whose value
represents a group. In such cases, the collective is also a decision
procedure.

The CUDA synchronization mask and the represented authority set are distinct
objects. The \texttt{*\_sync} mask constrains which non-exited lanes named by
the mask must execute the primitive for CUDA-defined behavior. The security
question is different: whether the collective decision represents the
membership, contributions, roles, and checked state authorized by the
application contract. The core attacks keep CUDA calls conforming and corrupt
the program-visible metadata that gives the collective its security meaning.
Accordingly, the evaluation and claims are limited to NVIDIA CUDA.

Participation-sensitive decisions also occur beyond simple warp votes. A
shuffle depends on a source lane and often a subwarp width. A scan or
reduction depends on ranks and contribution sets. A warp-aggregated operation
may elect one lane to represent the group. Barriers, shared-memory protocols,
and Cooperative Groups depend on the right threads reaching synchronization
and consuming shared state as the intended group
~\cite{nvidiaCooperativeGroups}. NVIDIA guidance treats participation mismatch
as a programming concern, especially under independent thread scheduling
~\cite{nvidiaCudaGuide,linCudaWarpPrimitives}. The same objects become
adversarial non-control data when a collective authorizes a security decision.

\subsection{Participation Metadata as Decision-Making Data}

Participation-sensitive state is ordinary CUDA programming state. Libraries
and applications use cooperative primitives, scans, reductions, compaction,
tiled dense kernels, graph frontiers, and top-\(k\) or selection kernels
~\cite{blelloch1990prefix,sengupta2007scan,sengupta2008scan,nvidiaCub,
nvidiaThrust,cudpp,moderngpu,merrill2016scan,kerr2017cutlass,
chetlur2014cudnn,wang2016gunrock,johnson2021faiss}. Benchmark suites also use
these patterns as heterogeneous workload kernels~\cite{che2009rodinia}. These
systems use masks, frontiers, row descriptors, valid flags, ranks, group
labels, widths, and leaders as performance and correctness metadata.

The same metadata becomes security-relevant when it defines the authority of a
collective decision. Participation metadata can be range-valid, in bounds, and
used by a CUDA-conforming call while still naming the wrong membership,
contribution set, role, or checked state. Collective Semantic Corruption
targets this gap: metadata that is valid as CUDA input is corrupted so that
the resulting collective decision represents unauthorized participation
authority.

The gap appears in common forms. A legal mask can exclude the lane holding a
rejecting record. An in-range source lane can name the wrong representative for
a shuffle or warp-aggregated commit. An in-bounds descriptor can carry a stale
count, epoch, or group label after validation. These are ordinary CUDA values
until a security-relevant collective treats them as authority.

\subsection{Limits of GPU Security}

Prior GPU security work establishes CUDA device code as an attack surface.
Recent exploitation work demonstrates memory corruption, code injection, ROP,
and code reuse in SASS-level device code~\cite{guo2024gpu,roels2025cuda}.
General memory-safety tools and compiler defenses target complementary
spatial and temporal memory objects
~\cite{serebryany2012asan,akritidis2009baggy,nagarakatte2009softbound,
nagarakatte2010cets}. GPU memory-safety systems such as GPUShield and CUSAFE
address important corruption causes through bounds checking or
memory-corruption detection~\cite{lee2022gpushield,lu2026cusafe}. Other work
studies GPU leakage and side channels
~\cite{naghibijouybari2018rendered,wang2024gpuzip,zhao2024owl}.

Those mechanisms primarily address causes of corruption or leakage channels.
CSC focuses on the collective use site: CUDA-resident metadata is syntactically
valid and consumed by a CUDA-conforming primitive, yet the decision uses
corrupted participation authority.

The resulting participation-authority contract names the site-local state that
determines which lanes are represented, what they contribute, who may speak for
the group, and which checked state reaches use.

%% file: sections/03_participation_integrity.tex
\section{Participation Authority}
\label{sec:model}

\subsection{Authority-Bearing Metadata}
At a CUDA collective decision, \emph{participation authority} is the relation
that determines which lanes are represented, what they contribute, which lane
may speak for the group, and which checked state reaches commit. The object is
security-critical when it defines the authority that the collective decision
is taken to represent. A corrupted scalar changes one value; corrupted
participation metadata changes which principals count.

This distinction is local to protected collective sites. A vote,
shuffle, reduction, barrier, cooperative group, or warp-aggregated commit is in
scope when its result authorizes, filters, routes, aggregates, synchronizes, or
commits data. We call such a site a \emph{protected collective site}. Stored
predicates can carry contribution authority; participant
metadata can carry membership authority; role metadata can decide who speaks or
commits for the group; reloaded descriptors can break the binding between
validation and use.

The four contract fields answer different authority questions. \emph{Membership}
asks who the decision represents. \emph{Contribution} asks what evidence those
participants are allowed to contribute. \emph{Role} asks which lane may speak,
broadcast, reserve, or commit for the group. \emph{Temporal binding} asks
whether the state checked before the collective is still the state authorized
at use. These dimensions can fail independently: a lane may be excluded from
membership, included with a poisoned contribution, replaced as the group
representative, or checked under one descriptor and committed under another.

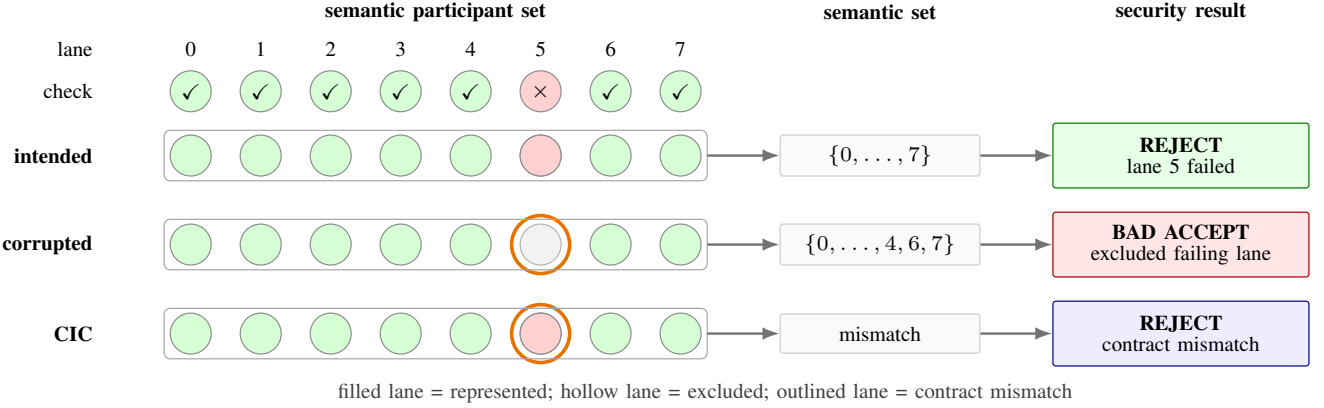
\begin{figure*}[t]
  \centering
  \resizebox{0.98\textwidth}{!}{%
  \begin{tikzpicture}[
    x=0.82cm,
    y=0.58cm,
    lane/.style={
      circle,
      draw=black!45,
      minimum size=0.48cm,
      inner sep=0pt,
      font=\scriptsize
    },
    inlane/.style={lane, fill=green!16},
    rejectlane/.style={lane, fill=red!18},
    outlane/.style={lane, fill=black!5, draw=black!30, text=black!45},
    rowlabel/.style={font=\scriptsize, anchor=east},
    title/.style={font=\scriptsize\bfseries, anchor=east},
    verdict/.style={
      rounded corners=1pt,
      draw=black!35,
      fill=black!3,
      font=\scriptsize,
      align=center,
      minimum width=3.0cm,
      minimum height=0.76cm,
      inner xsep=4pt,
      inner ysep=3pt
    },
    safeverdict/.style={verdict, draw=green!45!black, fill=green!10},
    badverdict/.style={verdict, draw=red!65!black, fill=red!10},
    cicverdict/.style={verdict, draw=blue!50!black, fill=blue!7},
    maskbox/.style={
      rounded corners=1pt,
      draw=black!20,
      fill=black!2,
      font=\scriptsize,
      align=center,
      minimum width=2.34cm,
      minimum height=0.46cm
    },
    warning/.style={draw=orange!90!black, very thick},
    arrow/.style={-{Latex[length=2mm]}, thick, draw=black!55}
  ]
    \node[font=\scriptsize\bfseries] at (3.5,0.75) {semantic participant set};
    \node[font=\scriptsize\bfseries] at (9.85,0.75) {semantic set};
    \node[font=\scriptsize\bfseries] at (14.15,0.75) {security result};

    \node[rowlabel] at (-1.25, 0) {lane};
    \foreach \i in {0,...,7} {
      \node[font=\scriptsize] at (\i,0) {\i};
    }

    \node[rowlabel] at (-1.25,-0.85) {check};
    \foreach \i in {0,1,2,3,4,6,7} {
      \node[inlane] at (\i,-0.85) {$\checkmark$};
    }
    \node[rejectlane] at (5,-0.85) {$\times$};

    \node[title] at (-1.25,-2.15) {intended};
    \foreach \i in {0,...,7} {
      \node[inlane] (a\i) at (\i,-2.15) {};
    }
    \node[rejectlane] at (5,-2.15) {};
    \draw[black!35, rounded corners=2pt] (-0.38,-2.66) rectangle (7.38,-1.64);
    \node[maskbox] (maska) at (9.85,-2.15) {$\{0,\ldots,7\}$};
    \node[safeverdict] (reject) at (14.15,-2.15)
      {\textbf{REJECT}\\[-1pt]lane 5 failed};
    \draw[arrow] (7.38,-2.15) -- (maska.west);
    \draw[arrow] (maska.east) -- (reject.west);

    \node[title] at (-1.25,-3.95) {corrupted};
    \foreach \i in {0,1,2,3,4,6,7} {
      \node[inlane] (b\i) at (\i,-3.95) {};
    }
    \node[outlane] (b5) at (5,-3.95) {};
    \draw[warning] (5,-3.95) circle[radius=0.34cm];
    \draw[black!35, rounded corners=2pt] (-0.38,-4.46) rectangle (7.38,-3.44);
    \node[maskbox] (maskb) at (9.85,-3.95) {$\{0,\ldots,4,6,7\}$};
    \node[badverdict] (accept) at (14.15,-3.95)
      {\textbf{BAD ACCEPT}\\[-1pt]excluded failing lane};
    \draw[arrow] (7.38,-3.95) -- (maskb.west);
    \draw[arrow] (maskb.east) -- (accept.west);

    \node[title] at (-1.25,-5.75) {\shortsys{}};
    \foreach \i in {0,...,7} {
      \node[inlane] (c\i) at (\i,-5.75) {};
    }
    \node[rejectlane] at (5,-5.75) {};
    \draw[warning] (5,-5.75) circle[radius=0.34cm];
    \draw[black!35, rounded corners=2pt] (-0.38,-6.26) rectangle (7.38,-5.24);
    \node[maskbox] (maskc) at (9.85,-5.75) {mismatch};
    \node[cicverdict] (cic) at (14.15,-5.75)
      {\textbf{REJECT}\\[-1pt]contract mismatch};
    \draw[arrow] (7.38,-5.75) -- (maskc.west);
    \draw[arrow] (maskc.east) -- (cic.west);

    \node[font=\scriptsize, text=black!75, align=center] at (7.35,-6.95)
      {filled lane = represented; hollow lane = excluded; outlined lane = contract mismatch};
  \end{tikzpicture}}
  \caption{A collective can reach the same program site while representing
  different lanes. Lane 5 fails the check, corrupted participation metadata
  excludes it, and \shortsys{} restores the trusted set. The masks shown are
  semantic participant sets.}
  \label{fig:participant-mask-example}
\end{figure*}

\subsection{Contract Objects}
A participation-authority contract names the fields that make a collective
decision meaningful: represented lanes, trusted mask or domain, allowed
contributions, valid roles, and validation-to-use state. It is derived before
the attack from an independent source available to the program or trusted
protocol. Table~\ref{tab:spi-objects} lists the contract objects,
representative trusted sources, and corrupted objects for a collective
operation $c$ at program point $\ell$.

\begin{table}[t]
  \centering
  \scriptsize
  \caption{Participation-authority contract fields and representative sources.}
  \label{tab:spi-objects}
  \setlength{\tabcolsep}{2pt}
  \resizebox{\columnwidth}{!}{%
  \begin{tabular}{@{}p{0.09\columnwidth}p{0.31\columnwidth}p{0.27\columnwidth}p{0.23\columnwidth}@{}}
    \toprule
    Object & Meaning & Trusted source & Corrupted object \\
    \midrule
    $I_\ell,M_\ell$ & represented lanes and trusted mask/domain & public size, launch shape, verified row domain & descriptor mask, group label \\
    $A_\ell$ & observed runtime set, not a trusted source & -- & loaded metadata-defined set \\
    $P_\ell$ & predicates, checks, values, or vote inputs & recomputed bounds or validity check & stored valid flag, include bit \\
    $R_\ell$ & rank, leader, source, width, tile, or group label & fixed leader or canonical rank rule & \texttt{leader\_lane}, source, width \\
    $\Phi_\ell$ & validation-to-use binding & frozen descriptor, checked copy, epoch & reloaded descriptor, stale count \\
    \bottomrule
  \end{tabular}
  }
\end{table}

The intended set $I_\ell$ is not required to be the full warp. It may be a tile, a tail fragment, a sparse-row group, a cooperative-group partition, or the lanes assigned to a batch item. Dynamic participation is normal in CUDA; it becomes a security problem when runtime metadata makes the collective represent a different authority set from the one the program intended.

Membership comes from public lane predicates such as
\texttt{lane < n} or verified row domains. Contribution comes from recomputing
validity or bounds predicates at authorization. Role comes from a canonical
rule, such as first valid lane, expected shuffle width, or active-lane rank.
Temporal binding comes from a frozen descriptor, checked copy, or epoch carried
to commit. These sources make the reference auditable: the contract is fixed
before the attack mode is selected and independent of the attacked output.

The \emph{trusted reference} is the protected observation obtained when the
collective consumes the independently derived contract fields rather than the
observed mutable metadata. Evaluation compares each attacked or hardened run
against this reference.

A trusted source is admissible only under five operational conditions. It is
\emph{pre-attack fixed}: the source is chosen before the observed metadata is
mutated. It follows an \emph{independent path}: the attacker cannot alter it
through the same metadata path consumed by the vulnerable collective. It is
\emph{available at the use site}: the kernel or trusted protocol can derive,
recompute, check, or freeze it before authorization. It is
\emph{policy-determining}: it defines the intended authority rather than being
recovered from the attacked output. Finally, it is \emph{auditable}: the
artifact records the trusted source, corrupted object, and binding relation.
The \shortsys{} site-local guarantee starts after this contract is constructed,
and dynamic policy changes must be reflected in the independently derived
contract before the collective uses them.

Our evaluation follows this rule directly. Tail validation derives
membership from public \texttt{lane < n}. Retrieval derives contribution
authority by recomputing candidate validity from candidate data. Shuffle-role
cases use a fixed leader policy. TOCTOU cases use a checked frozen descriptor.
Queue-rank cases derive output ownership from the canonical active-lane rank.

Figure~\ref{fig:participant-mask-example} illustrates the membership case. Each
lane validates one item and a warp vote accepts the group only when every
intended participant approves. The intended set $I_\ell$ is the lanes with
\texttt{lane < n}; the observed set $A_\ell$ is the set named by runtime
participation metadata. If metadata excludes a rejecting lane from the semantic
vote, then a CUDA-conforming \texttt{\_\_all\_sync} can return true. The kernel
reaches the intended collective site. The failure is the represented
participant set.

\subsection{Local Contract Criterion}
The trusted contract for a participation-sensitive operation is
$C_\ell=(I_\ell,M_\ell,P_\ell,R_\ell,\Phi_\ell)$. A contract builder
$K_\ell(\sigma)$ derives this tuple from independent state. The vulnerable
site consumes observed metadata $O_\ell(\sigma)$: values loaded, supplied, or
reloaded at the collective site.

A hardened use site relates these objects with a field-specific binding
$\mathsf{bind}_\ell(K_\ell(\sigma),O_\ell(\sigma))$. The binding may be mask
equality, predicate recomputation, role canonicalization, a frozen descriptor,
or an epoch check. The evaluated \shortsys{} wrappers usually enforce the
stronger condition that the consumed membership, contribution, role, and
temporal fields match the trusted contract fields.

The evaluation oracle compares protected observations. Let $O^c_\ell$ denote
the fields actually consumed by the collective. A correct execution satisfies
\[
  \mathsf{SecObs}(c(O^c_\ell)) =
  \mathsf{SecObs}(c(C_\ell)).
\]
Here \textsf{SecObs} is the kernel's protected observation: accept/reject,
threshold result, representative value, synchronization-safe state, or another
security decision named by the reference behavior. The contract is fixed before
evaluation; the equality scores executions against that contract.

A CSC instance is the corresponding protected-site violation:
\[
  \begin{aligned}
  \mathsf{CUDAConforming}(\ell) &\wedge
  \mathsf{RangeValid}(O_\ell)\\
  &{}\wedge\ \mathsf{SecObs}(c(O_\ell)) \ne
  \mathsf{SecObs}(c(C_\ell)).
  \end{aligned}
\]
The first conjunct keeps the call discipline in scope, the second separates CSC
from malformed metadata, and the inequality names the security failure.

\textbf{Site-local binding proposition.}
If every authority-bearing field consumed by a protected collective is derived
from $K_\ell(\sigma)$, checked against $K_\ell(\sigma)$, or frozen between
validation and use, then mutations of unbound observed metadata cannot change
\(\mathsf{SecObs}(c(C_\ell))\) through membership, contribution, role, or
temporal-binding authority at that site. This is a reference-monitor property
for the named site, not whole-program memory safety.

The distinction between $A_\ell$ and $M_\ell$ is intentional. $M_\ell$ is the
mask or domain value passed to a primitive or wrapper. $A_\ell$ is the set of
lanes whose values or synchronization actually affect the result. They usually
coincide, but they can diverge when lanes named by a mask do not contribute,
when predication removes lanes, when a cooperative-group object is built from
labels, or when a synchronization site has different reachability from the mask
used nearby.

The model identifies corrupted data that acts as participation authority,
distinct from ordinary operands. A protected collective
must consume fields that are independently derived, recomputed, explicitly
bound, or frozen with respect to $C_\ell$. Benign and attacked executions can
use the same kernel binary, reach the same collective site, and execute the same
CUDA collective while differing only in the authority represented by
$A_\ell$, $M_\ell$, $P_\ell$, $R_\ell$, or $\Phi_\ell$.

\subsection{Authority-Corruption Dimensions}
A Collective Semantic Corruption instance occurs when the instruction stream
remains legal, but one contract object no longer matches the intended
computation. The program can execute the expected basic blocks and call
the expected CUDA intrinsic. The violation is in who participates, what those
lanes contribute, which lane speaks, or which checked state reaches commit.

Dynamic masks, divergent branches, and partial tiles are normal CUDA idioms
when they follow the intended contract. A site is security-relevant when the
collective result authorizes, filters, aggregates, synchronizes, routes, or
commits data under an intended participant contract.
Section~\ref{sec:threat-model} gives the operational selection rule used by
the evaluation.

\begin{tcolorbox}[
  colback=black!2,
  colframe=black!35,
  boxrule=0.35pt,
  arc=1pt,
  left=4pt,
  right=4pt,
  top=3pt,
  bottom=3pt,
  fontupper=\small
]
\textbf{Defined-behavior core.} In the core attacks, all lanes required by the
CUDA synchronization mask execute the same collective at the same program
point. The corrupted object is application metadata: participant predicates,
contributions, roles, descriptors, or validation-to-use state. Hardware active
masks, reconvergence state, and \texttt{*\_sync} participation remain
conforming in the core set; synchronization-divergence cases are reported
separately.
\end{tcolorbox}

We call the concrete family of such attacks \emph{Collective Semantic
Corruption}. These are non-control-data attacks: the corrupted object is
program-visible semantic state that affects a warp, tile, block, or cooperative
group. Hardware active masks, reconvergence stacks, and scheduler state are
not the corrupted object.

The criterion applies when program-visible metadata defines who counts, what
counts, who speaks, or which checked state reaches commit. Memory safety can
prevent some corruptions, and control-flow integrity can validate branch and
call targets, but neither names the participation authority a CUDA collective
is supposed to represent.

%% file: sections/04_threat_model.tex
\section{Threat Model}
\label{sec:threat-model}

\subsection{Threat Setting and Site Selection}

We consider CUDA kernels in which a collective result controls a protected
decision: accepting an input, authorizing a write, filtering data, selecting a
representative lane, aggregating evidence, routing work, or committing state.
The collective may be a vote, ballot, shuffle, reduction, scan, barrier,
cooperative group, or warp-aggregated commit.

A collective site is counted as an in-scope CSC instance only when three
conditions hold before evaluation. First, the collective output controls a
security-relevant action, such as accept/reject, route/drop, write/commit,
representative selection, threshold admission, or synchronization-safe
consumption. Second, the site consumes participation metadata reachable from
untrusted input, stale state, prior-stage output, or a memory-corruption
trigger. Third, the intended participation contract is independently derivable
from public shape, launch geometry, canonical tile policy, recomputed
predicates, checked descriptors, frozen copies, epochs, or fixed role rules.
Other sites remain review targets or controls.

The trusted code context is the kernel binary, launch configuration, CUDA
runtime, and GPU hardware. The attacker acts through program-visible data
consumed by device code. The relevant authority boundary is crossed when
untrusted, stale, or corrupted metadata becomes participation authority for a
collective decision. Appendix Table~\ref{tab:trusted-vs-untrusted} gives
representative trusted-source and untrusted-metadata pairs.

\subsection{Attacker Capability and Channels}

The attacker can trigger the target kernel with chosen inputs and observe the
resulting service decision, output, error, or timeout. The attacker can affect
participation-defining data through three channels. The channels differ in
cause; all make a collective consume participation authority that differs from
the independently derived contract.

For adversarial metadata, a host request can carry a syntactically valid
descriptor, count, label, or leader field that the kernel later treats as
authority. For stale state, one kernel or phase can validate metadata while a
later collective reloads a changed descriptor, epoch, or count. For a
memory-corruption trigger, a device-side overwrite can reach a valid flag,
mask, rank, or descriptor before the collective consumes it.

\begin{table}[t]
  \centering
  \scriptsize
  \caption{Attacker channels and assumptions. The shared consequence is
  corrupted participation authority at a collective use site.}
  \label{tab:threat-channels}
  \setlength{\tabcolsep}{2pt}
  \resizebox{\columnwidth}{!}{%
  \begin{tabular}{@{}p{0.17\columnwidth}p{0.24\columnwidth}p{0.22\columnwidth}p{0.20\columnwidth}p{0.21\columnwidth}@{}}
    \toprule
    Channel & Attacker capability & Needed precondition & Realistic source & Not assumed \\
    \midrule
    Adversarial metadata &
    choose well-formed descriptors, counts, labels, or leader indices &
    kernel treats metadata as authority &
    request descriptors, batch metadata, routing labels &
    PC control or malformed CUDA calls \\

    Stale/logical state &
    cause checked state to differ from later consumed state &
    check/use drift across reload, phase, or kernel boundary &
    prior-stage output, descriptor reuse, epoch drift &
    arbitrary device-memory overwrite \\

    Memory-corruption trigger &
    corrupt a participation object through a device-side write &
    bug reaches mask, predicate, role, or descriptor storage &
    overflow, intra-object overwrite, stale pointer &
    code injection, ROP, or arbitrary writes to all memory \\
    \bottomrule
  \end{tabular}
  }
\end{table}

Adversarial metadata is syntactically valid input that the kernel later treats
as authority. Stale or logical state remains in bounds but no longer represents
the state that was validated. Memory-corruption triggers are device-side writes
into program-visible participation state before a collective observes it. The
evaluation reports the channels separately to avoid conflating input
validation, stale state, and memory safety.

Representative dataflows include batch validation, retrieval admission, graph
frontiers, multi-kernel descriptor pipelines, and memory-corruption overlays.
Across those settings, the attacker-influenced object may be a valid mask,
membership predicate, lane rank, output slot, epoch, count, or role value; the
security effect is an invalid batch, candidate, frontier, or stale state being
accepted under the wrong participation authority.

\subsection{Scope}

The trusted contract is the intended participant set, mask or domain, per-lane
inputs, role metadata, and validation-to-use binding for a security-relevant
collective. The untrusted side is runtime data that can define, mutate, or
replace those objects. Runtime data may be used safely after the kernel binds
it to the contract through validation, freezing, recomputation, or an explicit
contract check.

The model excludes GPU program-counter control, device-code injection, ROP or
code reuse, instruction-stream patching, kernel replacement, driver compromise,
and direct writes to undocumented scheduler, active-mask, or reconvergence
hardware state. Some cases may change data predicates that guard branches or
commits while code, branch targets, and gadget chains remain fixed. Each
attack needs a concrete path from chosen input, stale state, or a memory bug to
a participation-authority object; arbitrary writes to all device memory are
part of a strictly stronger memory-corruption model and are not required for
CSC.

CPU exploitation, GPU side channels, multi-tenant interference, and hardware
fault injection matter here only when they provide one of the stated channels.
Memory safety remains important because it can prevent some causes of
participation corruption. CSC studies the consequence after participation
metadata has become corrupted, stale, or attacker-controlled and is treated as
decision authority by a collective.

\subsection{Security Objectives}

The attacker aims to make a collective operation produce a security-relevant
result over invalid participation authority. Concrete outcomes include invalid
input accepted, a rejecting lane excluded from consensus, a wrong lane
broadcast as a leader, a poisoned aggregate, stale shared state consumed after
partial synchronization, or a commit performed under mutated participation
metadata. Denial of service is in scope for barrier and synchronization cases,
but the primary target is a silent integrity failure: the kernel finishes
without a CUDA error, timeout, trap, or visible control-flow anomaly, while the
decision or output violates the intended contract.

The target programs are CUDA kernels where a collective result authorizes,
filters, aggregates, routes, or commits data: validation kernels, sparse gather
and scatter, reductions, scans, graph-frontier processing, embedding
similarity, top-\(k\) and admission pipelines, tiled shared-memory kernels,
cooperative-group code, and warp-aggregated atomics. These programs are in
scope when participation metadata has a concrete path from one of the attacker
channels above to the collective use site.

%% file: sections/05_attacks.tex
\section{Collective Semantic Corruption}
\label{sec:attacks}

\subsection{Common Pattern and CUDA Boundary}
Collective Semantic Corruption has one pattern: a kernel reaches the intended
collective site, but program-visible metadata supplies the wrong participation
authority. The metadata may be adversarial, stale, or memory-corrupted. The
CUDA primitive can remain conforming while the collective counts the wrong
members, predicates, roles, or checked state.

We organize CSC by the contract field controlled by mutable metadata:
\emph{membership} (who is represented), \emph{contribution} (what evidence
counts), \emph{role} (who speaks or commits), and \emph{temporal binding}
(which checked state reaches use). Labels such as mask substitution,
predicate poisoning, vote spoofing, leader confusion, barrier drift, and
collective TOCTOU are concrete instances of these fields, not separate attack
classes.

Unless otherwise stated, examples use the pseudocode variable
\texttt{call\_mask} for a CUDA-conforming call mask. Participation is modeled
as a predicate input to the collective; this is membership-authority
corruption when metadata changes the semantic participant set represented by
the vote. Evaluated kernels implement the core cases as straight-line
full-warp calls or with branches that ensure the named lanes execute.
Malformed masks and synchronization-sensitive cases are separated as
synchronization-sensitive results.

A dynamic mask or partial tile is a normal CUDA idiom when it follows the
intended contract. It becomes an in-scope CSC instance when reachable metadata
defines membership, contribution, role, or temporal state for a protected
decision and the resulting collective produces a mismatch with the trusted
reference.

\subsection{Membership-Authority Corruption}
Membership-authority attacks change who the collective represents. The corrupted
object can be a mask, group label, tail predicate, tile descriptor, or
barrier predicate. The collective then counts the wrong authority set while
keeping the CUDA synchronization mask conforming.

\begin{codeatom}[12]
uint32_t call = call_mask;
bool trusted_member = lane < n;
bool observed_member =
    (desc->participant_mask >> lane) & 1u;
bool ok =
    !trusted_member || validate_item(x[lane]);
// Vulnerable: metadata defines membership.
bool vote_input =
    !observed_member || ok;
bool group_ok =
    __all_sync(call, vote_input);
if (lane == leader && group_ok) {
    accept();
}
\end{codeatom}

If \texttt{desc->participant\_mask} drops a lane whose \texttt{ok} is false,
the vote can still accept based on the remaining metadata-selected lanes.
The CUDA call mask defines who executes; the unsafe mask defines who is
counted. A mask can use only valid lane bits and still violate the contract.
The defense therefore binds membership to a trusted source such as
\texttt{lane < n}, a verified sparse descriptor, or a frozen tile contract.
This same issue covers corrupted group labels and phase predicates; barrier
hangs and timeouts are secondary.

\subsection{Contribution-Authority Corruption}
Contribution attacks keep the collective site and usually the membership,
but change which lanes contribute. The corrupted object is a validity flag,
bounds predicate, override bit, contribution value, or inclusion predicate.

\begin{codeatom}[8]
bool trusted_member = lane < n;
uint32_t call = call_mask;
// Vulnerable: stored validity drives the vote.
bool ok = !trusted_member ||
    (metadata_valid[lane] && bounds_flag[lane]);
bool vote_input = !trusted_member || ok;
if (__all_sync(call, vote_input)) {
    publish_batch();
}
\end{codeatom}

The attack changes \texttt{metadata\_valid} or \texttt{bounds\_flag} so that a
lane contributes approval when trusted validation would reject it. Reductions
use the same authority bug:

\begin{codeatom}[10]
int include = valid[lane];
float contrib = include ? score[lane] : 0.0f;
float sum = warp_reduce_sum(contrib);
int count = warp_reduce_sum(include);
bool accept_group =
    count > 0 && sum / count < threshold;

if (lane == canonical_committer() &&
    accept_group) {
    accept();
}
\end{codeatom}

The reduction opcode stays fixed; the corrupted field changes which evidence
the reduction represents. The effect is group-level: the corrupted predicate
becomes an authorization result, aggregate, or threshold decision.

\subsection{Role-Authority Corruption}
Role attacks corrupt who is allowed to represent the group. The corrupted
object can be a source lane, leader ID, rank, shuffle width, output owner, or
warp-aggregated commit leader.

\begin{codeatom}[13]
bool trusted_member = lane < n;
uint32_t call = call_mask;
uint32_t expected = __ballot_sync(call, trusted_member);

// Vulnerable: role comes from mutable metadata.
// Active lane, wrong role.
int src = desc->leader_lane;
int representative =
    __shfl_sync(call, local_value, src, 32);

if (lane == canonical_committer(expected))
    commit_representative(representative);
\end{codeatom}

A malicious \texttt{src} can be in range and semantically wrong. The
program may have intended the first valid lane, the lane with the minimum
rank, or a policy-defined leader. If metadata chooses a different source, a
legal shuffle broadcasts the wrong value. Role attacks cover shuffle-helper
patterns, warp-shuffle domains, warp-aggregated atomics, and leader-based
admission or output-slot commits.

\subsection{Temporal-Binding Corruption}
Temporal attacks corrupt the binding between validation and later collective
use. The kernel checks a descriptor, mask, predicate, or phase, but later
reloads mutable state or consumes state that another kernel has changed.

\begin{codeatom}[13]
uint32_t checked = desc->mask;
if (checked != expected_mask)
    trap();

// Later code reloads mutable metadata.
uint32_t used = desc->mask;
bool observed_member = (used >> lane) & 1u;
uint32_t call = call_mask;
bool vote_input =
    !observed_member || ok;
if (__all_sync(call, vote_input)) {
    commit();
}
\end{codeatom}

The check can be correct and fail to protect the commit. The
participation-authority contract was validated for \texttt{checked}; the
collective consumes \texttt{used}. Temporal attacks include single-kernel
TOCTOU, stale shared or global state, and multi-kernel pipelines that treat
previously validated metadata as authoritative.

\subsection{Implications}
CSC targets the participation authority consumed by a CUDA collective decision.
Membership attacks change who counts; contribution attacks change what counts
as approval or evidence; role attacks change who speaks or commits for the
group; temporal attacks change which checked state reaches use. The evaluation
instantiates these dimensions with vote masks, sparse
gather domains, grouped reductions, validity predicates, admission guards,
shuffles, scan widths, and multi-kernel metadata chains.

The same boundary explains the negative controls. Unused corrupted metadata
preserves the reference decision, and recomputing the trusted predicate before
authorization removes the corrupted field from the authority path. Range
validation alone is insufficient for the same reason: a source lane can be
between 0 and 31, a mask can contain only legal bits, and a count can be within
buffer bounds while naming the wrong authority set. The contract relation must
be checked alongside the value range.

%% file: sections/06_evaluation.tex
\section{Empirical Evidence}
\label{sec:evaluation}

\subsection{Evidence Blocks and Scope}
The evaluation uses the 102 CUDA-defined instances as a
contract-conformance suite, not a prevalence estimate. It asks whether each
authority dimension can produce a trusted-reference mismatch while preserving
the CUDA-defined core boundary, whether different channels can reach the same
authority objects, and whether \shortsys{} preserves the trusted reference
when those fields are bound. Supporting blocks cover fuzzed state spaces,
service-visible impact, tool boundaries, review surface, and cost.

We run the evaluation on an NVIDIA GeForce RTX 4070 Laptop GPU. Each execution
is checked against a trusted reference: collective output, application
decision, invalid-acceptance status, or bounded synchronization outcome. The
execution modes are \emph{benign}, \emph{attack}, \emph{hardened}, and
\emph{control}. A \emph{case} is a kernel scenario; a \emph{run} is one mode
for that case; an \emph{evaluation instance} is a counted observation used for
a reported result. Table~\ref{tab:evaluation-claims} gives the evidence
hierarchy, and Appendix~\ref{app:run-accounting} gives the full traceability.

\begin{table}[!t]
  \centering
  \scriptsize
  \caption{Evaluation hierarchy. Support blocks answer narrower questions than
  the core contract-conformance suite and are not additive.}
  \label{tab:evaluation-claims}
  \setlength{\tabcolsep}{3pt}
  \resizebox{\columnwidth}{!}{%
  \begin{tabular}{@{}p{0.27\columnwidth}p{0.30\columnwidth}p{0.36\columnwidth}@{}}
    \toprule
    Evidence block & Main result & What it supports \\
    \midrule
    Core contract suite & 102/102 trusted-reference mismatches; 102/102 hardened preserved & CUDA-defined membership, contribution, role, and temporal failures \\
    Sync boundary & 13 secondary instances & synchronization-sensitive cases stay outside the core claim \\
    Fuzzing & 65,536 model states; 65,536 CUDA states & contract model is not tied to one manual example \\
    Service harnesses & 12/16 invalid retrieval accepts become 0/16 & service-visible impact and negative controls \\
    Site/tool boundary & 59/59 paired hashes; 21 sanitizer checks & same-site evidence and semantic-tool boundary \\
    Cost and ablation & 51 timings; 133 field observations & \shortsys{} cost and field-coverage limits \\
    \bottomrule
  \end{tabular}
  }
\end{table}

Two counted sets define the main boundary: 102 CUDA-defined core instances and
13 synchronization-sensitive instances reported separately. Other counts are
support views over the same artifact and should not be added to the 102 core
instances. Appendix Tables~\ref{tab:appendix-evaluation-accounting}
and~\ref{tab:core-instance-breakdown} give the secondary accounting and grouped
breakdown.

\subsection{All Four Authority Dimensions Produce Mismatches}
We first isolate the four contract fields in compact CUDA kernels. Each kernel
keeps the CUDA primitive and kernel path fixed, and changes only the metadata
field that carries one participation-authority dimension. The canonical
patterns are membership through a validation mask, contribution through an
outlier-reduction predicate, role through a shuffle source or leader, and
temporal binding through vote-to-commit descriptor drift.

The primitive-isolation set yields 35/35 attack-mode trusted-reference
mismatches. The fuzzers stress the same contract model rather than adding to
this denominator: the model fuzzer uses seed 424242, a 20\% control rate, and
an 8-lane contract model; the CUDA differential fuzzer uses seed 1337, one
full-mask warp per state, and 16,384 states per dimension. Both generate bad
lanes, source lanes, control status, and checked/reloaded state, then compare
vulnerable and hardened observations with the trusted reference.

Each fuzzer checks 65,536 generated states. The summaries report 52,340 model
divergences and 52,622 CUDA divergences, with 0 \shortsys{} false negatives
and 0 control false positives. A scalar-vs-contract control isolates the
participation-authority component: scalar-only corruption preserves the
decision in 1/1 run, while membership, contribution, role, and temporal
authority corruptions flip it in 4/4 runs; hardened variants preserve and
detect in 4/4 runs. Appendix Table~\ref{tab:fuzzing-methodology} gives the
generator rules, seeds, valid-state counts, controls, and per-dimension
divergences.

\subsection{Different Trigger Channels Reach the Same Authority Object}
We next separate causes from consequences. The same participation-authority
failure is triggered through three attacker channels. Adversarial metadata
models syntactically valid descriptors, masks, labels, counts, or leader values
that become participation authority. Stale or in-bounds state models
validation-to-use drift, descriptor reuse, and reloaded mutable metadata.
Device-side memory corruption models an out-of-bounds or intra-object write
that reaches program-visible participation state before the collective uses it.

The memory-corruption overlay replaces direct metadata mutation with a
device-side overflow trigger. Across all 25 benchmark cases, 55/55
overflow-triggered attack runs execute the trigger and produce a mismatch with
the trusted reference. Device memory bugs can reach the same
participation consequence, while stale-state and adversarial-metadata cases
cover paths without an out-of-bounds write at the use site.

\subsection{Retrieval Admission Shows Service-Visible Impact}
The clearest end-to-end application case is the retrieval admission guard
summarized in Table~\ref{tab:retrieval-case-study}. A host request reaches a
prior-stage validity mask, a warp collective scores effective lanes, and a
threshold admits or rejects the candidate. The trusted source is recomputed
candidate validity at the collective site; the corrupted object is the stored
validity mask. The CUDA call remains conforming, but the vulnerable consumer
lets the stored mask include lane 7 even though the trusted contract excludes
it. \shortsys{} recomputes validity, derives the effective mask from that
trusted predicate, and reports a mismatch when the stored mask disagrees.

\begin{table}[t]
  \centering
  \scriptsize
  \caption{Retrieval/admission case and service result.}
  \label{tab:retrieval-case-study}
  \label{tab:service-expanded}
  \setlength{\tabcolsep}{2pt}
  \resizebox{\columnwidth}{!}{%
  \begin{tabular}{@{}p{0.42\columnwidth}p{0.23\columnwidth}p{0.25\columnwidth}@{}}
    \toprule
    Evidence & Vulnerable & \shortsys{} / trusted \\
    \midrule
    Effective participant set & lane 7 included & lane 7 excluded \\
    Bad-candidate score / decision & 2.48 / admit & 0.50 / reject \\
    Attackable invalid requests & 12/16 invalid accept & 0/16 invalid accept \\
    Non-flipping attackable requests & 4 reference preserved & 4 reject + detect \\
    Benign and non-triggering controls & 16 preserved & 16 preserved; 0 FP \\
    Total batch & 12 invalid accepts & 0 invalid accepts \\
    \bottomrule
  \end{tabular}
  }
\end{table}

The service harness runs 32 deterministic requests covering
adversarial metadata, stale prior-stage metadata, memory-overlay attacks,
benign controls, non-triggering attack controls, and unused-metadata controls.
Without \shortsys{}, 12/16 attackable invalid requests become invalid accepts.
With \shortsys{}, all 32 requests preserve the trusted security outcome:
16 attackable invalid requests raise mismatch detections, and 16 benign or
non-triggering controls preserve the reference with 0 false-positive
detections.

\subsection{Compact Workloads Reuse the Same Authority Object}
The remaining experiments check whether the same authority objects appear
beyond one guard. Validation reductions produce consensus bypass, outlier
filters poison aggregates, and sparse gather, shared-state, and multi-kernel
cases affect data movement and checked state. Reduced idiom bridges cover
shuffle helpers, warp reductions, tiled-validity metadata, CUDA sample-style
vote/reduction, warp-shuffle domains, grouped reductions, and queue
reservation. The bridge names indicate idiom inspiration only; they are not
vulnerability claims against the named libraries.

The compact application-level guards provide the most direct security-impact
evidence. Embedding-style admission, retrieval/admission, and anomaly detection
produce six trusted-reference mismatches, six invalid acceptances, and six
decision flips. Batchnorm, packet, block-level frontier, multi-kernel temporal,
and queue-rank guards add ten instances that admit an invalid sample, packet,
frontier, or stale descriptor state; hardened variants detect the mismatch and
preserve rejection. These instances enter the defined-behavior core as
membership, temporal, or role failures.

The queue-reservation frontier harness is an example of the CSC pattern in a
graph-processing idiom, not a Gunrock vulnerability claim. The pattern is
simple: a warp reserves a contiguous output window, and each lane writes its
frontier item at \texttt{base + rank}. The trusted contract says which
reservation base and lane ranks belong to the frontier window being checked.
In the attack, the invalid lane still writes to an in-range queue slot, but
corrupted base/rank metadata moves that write outside the consumed window. The
subsequent frontier check sees only valid entries and admits the frontier.

We include an API-shape check that calls Gunrock's \texttt{Queue::next()} to
exercise a real queue-reservation shape. The vulnerable and range-check-only
variants admit the invalid frontier in 6/6 instances because the corrupted
slot is still inside queue capacity. With \shortsys{}, the
reservation-base/rank mismatch is detected and the frontier is rejected in 6/6
instances; one unused-state control preserves the reference.

\subsection{Controls and Tool Checks}
In the defined-behavior evaluation, attack instances produce 102/102
trusted-reference mismatches and hardened variants preserve the trusted
reference in 102/102 instances. The 13 synchronization-sensitive instances are
reported separately, and 0 UB-sensitive instances are used for the core result.
The evaluated attack cases include 38 invalid acceptances and 16 decision
flips; these are security-observation labels, not additional denominators.

Core instances keep the CUDA call discipline fixed: all lanes named by the
call mask execute the same collective at the same program point, while
range-valid metadata changes a semantic predicate, contribution, role, or
validated descriptor. Range checks, bounds checks, Compute Sanitizer, and
control-flow checks can show syntactic validity, lack of reported memory,
race, or synchronization errors, and expected site reachability; they do not decide whether
membership, contribution, role, and epoch match the trusted contract.

Negative controls check specificity. The evaluation includes 12 explicit
controls, and the retrieval grid adds 144/144 non-triggering controls; the
288/288 expected attack configurations flip the admission decision. The
representative Compute Sanitizer runs~\cite{nvidiaComputeSanitizer} report no
memory, race, or synchronization errors in 21 checks covering seven instances,
while those executions produce a trusted-reference mismatch.

Same-site paired checks match direct-trigger attacks with benign counterparts
for the same case, trigger class, and collective site. In the current paired
core replay, benign twins pass in 59/59 cases, attack twins fail in 59/59
cases, and SASS/PTX hash evidence is available for 59/59 pairs. A separate
marker harness records identical marker sequences for one membership,
contribution, role, and temporal pair each (4/4), while the attack decision
diverges and hardened mode reports a mismatch. These checks support the
same-site boundary; they are not full trace-equivalence evidence.
Appendix~\ref{app:same-site-protocol} gives the checklist.

\subsection{The Checker Identifies Candidate Review Sites}
\label{sec:review-surface-scan}

A lightweight checker estimates review surface in public CUDA code. Findings
are separate from the 102-instance attack claim: they mark code where masks,
predicates, shuffle roles, synchronization context, or memory-derived
predicates may influence a collective decision. A checker finding is a review
prompt, not a vulnerability report. Vulnerability evidence would also require
reachability, attacker influence, an independently derivable contract source,
a security-relevant decision, and a concrete trusted-reference mismatch.

Across eight CUDA repositories, the checker scanned 1,946 CUDA/C++,
PTX, and SASS-related files and reported 264 candidate sites. In a deterministic
manual audit of 100 findings, 45/61 labeled in-scope findings matched the
participation-authority pattern. Table~\ref{tab:checker-funnel} gives the
funnel and separates this manual-audit match rate from checker performance,
defect rate, or vulnerability prevalence.

\begin{table}[t]
  \centering
  \scriptsize
  \caption{Checker triage funnel. Counts describe review surface, not
  vulnerability prevalence.}
  \label{tab:checker-funnel}
  \setlength{\tabcolsep}{3pt}
  \resizebox{\columnwidth}{!}{%
  \begin{tabular}{@{}p{0.44\columnwidth}rp{0.37\columnwidth}@{}}
    \toprule
    Stage & Count & Interpretation \\
    \midrule
    CUDA/C++/PTX/SASS files scanned & 1,946 & search corpus \\
    Candidate collective sites & 264 & review prompts \\
    Manually audited findings & 100 & deterministic audit sample \\
    Labeled in-scope findings & 61 & participation-sensitive labels after manual audit \\
    Direct pattern matches & 45 & participation-authority pattern matches \\
    Benign pattern false positives & 16 & in-scope shape without trusted-reference mismatch \\
    Out-of-scope / deferred & 38 / 1 & outside CSC scope or postponed \\
    API-reachability follow-up & 1 & handled separately under disclosure \\
    \bottomrule
  \end{tabular}
  }
\end{table}

\subsection{Public-API Disclosure Follow-Up}
One checker row led to a separate API-reachability follow-up in a third-party
CUDA project. We treat it separately from the 102-instance suite and reduced
idiom bridges. The idiom was role/domain authority: a public configuration
value reached a warp shuffle as domain metadata, the public API accepted the
value, and the protected output differed from a slow reference. The intended
contract source is the API-level domain policy for the optimized warp path; the
corrupted object is the caller-controlled domain value consumed by the shuffle
helper. This follow-up shows public-input reachability for authority-bearing
collective metadata, but it is not part of the core contract-conformance suite.
The affected code path is under coordinated disclosure.

\subsection{Summary}
Overall, participation-authority corruptions are reproducible across the
contract suite, reachable through representative attacker channels, and visible
in compact security-decision harnesses. The CUDA-defined core establishes the
main limit: the collective site and primitive stay fixed while mutable
membership, contribution, role, or temporal metadata produces a
trusted-reference mismatch. The trigger-channel, retrieval, tool-boundary, and
checker results provide supporting evidence rather than additive denominators.

%% file: sections/07_binding_participation_authority.tex
\section{Binding Participation Authority}
\label{sec:pic}

\subsection{Design Principle}
\sys{} (\shortsys{}) is a site-local binding discipline for CUDA collective
non-control data. A hardened site names the participation-authority contract,
then makes the collective consume trusted or bound state. If runtime metadata
disagrees with the contract, hardened code rejects, reports, or falls back.

\textbf{\shortsys{} invariant.} No protected collective may consume
mutable participation metadata as authority unless that metadata has been
derived from, checked against, or frozen with respect to an independently
derived participation-authority contract at the collective site. If the
contract cannot be named, \shortsys{} cannot infer the policy.

\textbf{Site-local guarantee.} For a collective site $\ell$, if every
authority-bearing field consumed by the collective is derived from
$K_\ell(\sigma)$, checked against $K_\ell(\sigma)$, or frozen between
validation and use, then later corruption of unbound observed metadata
$O_\ell$ cannot change $\mathsf{SecObs}(c(C_\ell))$ through membership,
contribution, role, or temporal-binding authority at that site. The guarantee
is site-local: it assumes the trusted contract source is correct when
constructed and that the relevant authority fields have been named.

\begin{figure*}[t]
  \centering
  \resizebox{0.96\textwidth}{!}{%
  \begin{tikzpicture}[
    x=1cm,
    y=0.90cm,
    box/.style={
      rounded corners=2pt,
      draw=black!35,
      align=center,
      font=\scriptsize,
      minimum height=0.78cm,
      inner xsep=6pt,
      inner ysep=5pt
    },
    header/.style={box, fill=black!5, font=\scriptsize\bfseries,
      minimum width=4.20cm, minimum height=0.72cm},
    state/.style={box, fill=black!2, minimum width=4.20cm},
    check/.style={box, fill=black!6, minimum width=2.80cm},
    formula/.style={box, fill=black!3, minimum width=4.05cm,
      minimum height=0.76cm},
    pass/.style={box, draw=green!45!black, fill=green!8,
      minimum width=2.75cm, minimum height=0.70cm},
    fail/.style={box, draw=red!55!black, fill=red!7,
      minimum width=2.75cm, minimum height=0.70cm},
    arrow/.style={-{Latex[length=1.8mm]}, thick, draw=black!45},
    passarrow/.style={-{Latex[length=1.8mm]}, thick, draw=green!45!black},
    failarrow/.style={-{Latex[length=1.8mm]}, thick, draw=red!55!black},
    title/.style={font=\scriptsize\bfseries, align=center},
    note/.style={font=\scriptsize, align=center, text=black!70}
  ]
    \node[title] at (5.25,1.50) {\shortsys{} binding relation at collective site $\ell$};

    \node[header] (chead) at (0,0.42)
      {\textbf{Trusted contract $C_\ell$}\\computed / recomputed / frozen};
    \node[header, minimum width=2.80cm] (bhead) at (5.25,0.42)
      {\textbf{Binding}\\relation};
    \node[header] (ohead) at (10.50,0.42)
      {\textbf{Observed runtime state $O_\ell$}\\loaded / supplied / reloaded};

    \node[state] (c1) at (0,-0.85)
      {$I_\ell,M_\ell$: intended participants\\trusted mask/domain};
    \node[check] (b1) at (5.25,-0.85)
      {\textbf{membership}\\\texttt{mem\_ok}};
    \node[state] (o1) at (10.50,-0.85)
      {$A_\ell$: observed participants\\loaded mask / group / tile};

    \node[state] (c2) at (0,-2.05)
      {$P_\ell$: trusted predicates\\recomputed values / validity};
    \node[check] (b2) at (5.25,-2.05)
      {\textbf{contribution}\\\texttt{contrib\_ok}};
    \node[state] (o2) at (10.50,-2.05)
      {$P'_\ell$: stored predicates\\valid flags / contribution values};

    \node[state] (c3) at (0,-3.25)
      {$R_\ell$: canonical roles\\leader / source / rank / width};
    \node[check] (b3) at (5.25,-3.25)
      {\textbf{role}\\\texttt{role\_ok}};
    \node[state] (o3) at (10.50,-3.25)
      {$R'_\ell$: runtime roles\\descriptor leader / source lane};

    \node[state] (c4) at (0,-4.45)
      {$\Phi_\ell$: validated temporal state\\frozen descriptor / epoch};
    \node[check] (b4) at (5.25,-4.45)
      {\textbf{temporal}\\\texttt{time\_ok}};
    \node[state] (o4) at (10.50,-4.45)
      {$\Phi'_\ell$: current temporal state\\reloaded descriptor / phase};

    \foreach \l/\m/\r in {c1/b1/o1,c2/b2/o2,c3/b3/o3,c4/b4/o4} {
      \draw[arrow] (\l.east) -- (\m.west);
      \draw[arrow] (\r.west) -- (\m.east);
    }

    \node[formula] (formula) at (5.25,-5.78)
      {$\begin{aligned}
        \mathrm{bind}_\ell={}&
        \texttt{mem\_ok}\wedge\texttt{contrib\_ok}\\
        &{}\wedge\texttt{role\_ok}\wedge\texttt{time\_ok}
      \end{aligned}$};
    \draw[arrow] (b1.south) -- (b2.north);
    \draw[arrow] (b2.south) -- (b3.north);
    \draw[arrow] (b3.south) -- (b4.north);
    \draw[arrow] (b4.south) -- (formula.north);

    \node[pass] (pass) at (0.95,-5.78)
      {\textbf{true}\\use trusted/bound state};
    \node[fail] (fail) at (9.55,-5.78)
      {\textbf{false}\\reject/report/fallback};
    \draw[passarrow] (formula.west) -- (pass.east);
    \draw[failarrow] (formula.east) -- (fail.west);

    \node[note] at (5.25,-6.72)
      {The check is over contract fields. A range-valid mask, source lane, or
      descriptor must satisfy the binding relation.};
  \end{tikzpicture}}
  \caption{\shortsys{} binds CUDA collective non-control data to the
  participation-authority contract at the collective use site.}
  \label{fig:cic-contract-model}
\end{figure*}

\subsection{Binding Patterns}
Range validation checks syntax. Contract binding checks authority. The
retrieval guard shows the difference. The vulnerable path uses a prior-stage
mask as authority, so stale or corrupted metadata can change the voting set
without ever violating CUDA's call discipline. The hardened path recomputes
validity at the collective site, derives the expected mask from the trusted
contract, checks the stored mask against that expectation, and then reduces
only over the trusted participants. This binds the collective decision to the
contracted authority rather than to mutable observed metadata.

\begin{codetiny}
// Vulnerable: stored mask defines authority.
bool include = bit(desc->valid_mask, lane);
float sum = warp_reduce_sum(
    call_mask, include ? score[lane] : 0.0f);
// CIC: derive and check the authority first.
bool trusted = lane < n &&
    recompute_valid(candidate, lane);
uint32_t expected = cic_expected_mask(trusted);
cic_check_mask(expected, desc->valid_mask);
float safe_sum = cic_reduce_sum(
    call_mask, bit(expected, lane) ? score[lane] : 0.0f);
\end{codetiny}

\begin{table}[t]
  \centering
  \scriptsize
  \caption{\shortsys{} helpers bind concrete authority fields before use.}
  \label{tab:cic-operational-helpers}
  \setlength{\tabcolsep}{2pt}
  \resizebox{\columnwidth}{!}{%
  \begin{tabular}{@{}p{0.29\columnwidth}p{0.30\columnwidth}p{0.28\columnwidth}@{}}
    \toprule
    Helper & Contract input & Runtime action \\
    \midrule
    \texttt{cic\_expect\_mask} & public shape or verified domain & reject/report mask mismatch \\
    \texttt{cic\_all}, \texttt{cic\_any} & recomputed predicate & vote over trusted predicate \\
    \texttt{cic\_reduce} & trusted contribution predicate & zero untrusted contribution \\
    \texttt{cic\_shfl} & canonical source/rank/width & use canonical role metadata \\
    \texttt{cic\_commit} & frozen descriptor or epoch & reject stale commit state \\
    \bottomrule
  \end{tabular}
  }
\end{table}

The helper names are intentionally small. They form a contract-source
cookbook: public shape and launch geometry produce expected masks; recomputed
predicates produce trusted vote inputs; canonical role rules produce source
lanes, ranks, and widths; frozen descriptors and epochs bind validation to
commit. The wrapper can reject/report mismatches in production hardening, or log
expected and observed fields in debug mode.

\textbf{Validate membership and use trusted state.}
Unsafe code lets metadata decide which lanes the vote represents. Hardened code
builds an expected contract from the trusted membership predicate, compares
observed metadata against it, and uses the trusted predicate for the
collective.

\textbf{Recompute contributions at authorization.}
Stored valid bits, bounds flags, and vote inputs are safe only if they are
bound to the intended contract. \shortsys{} recomputes the predicate at the
authorization point or checks stored metadata against a trusted predicate.

\textbf{Bind roles and temporal state.}
Source lanes, widths, leaders, ranks, and group labels can be range-valid and
semantically wrong. \shortsys{} checks them against the role policy, such
as canonical first valid lane or expected shuffle width, then uses the
canonical role metadata. For TOCTOU cases, \shortsys{} freezes the validated
descriptor or binds it to the commit state so later reloads cannot silently
change the contract.

\subsection{Prototype}
The prototype API names the protected non-control object and routes the
collective through trusted or bound state. Appendix
Table~\ref{tab:cic-contracts} summarizes the compact API and field coverage.

The static checker flags dynamic masks, memory-derived masks, commit-edge
votes, dynamic shuffle source/width, sync under branch, and predicated stores.
Each finding marks a participation-sensitive site.

\shortsys{} can be used in three modes. Runtime hardening protects
security-sensitive kernels. Debug mode records expected and observed masks,
roles, epochs, protected decisions, and mismatch bits during testing. Static
review runs without runtime overhead and points developers to likely contract
sites.

\shortsys{} is a named-field discipline. The developer, checker, or hardening
helper must name the protected site and its membership, contribution, role, or
temporal policy. The guarantee starts after the trusted contract exists and
protects only the named fields. If an application changes policy dynamically,
the updated policy must be reflected in the trusted contract source before the
collective consumes the metadata.

Memory safety, isolation, taint tracking, and code review remain complementary:
they can prevent or identify causes of corruption. \shortsys{} addresses the
use-site question: whether program-visible metadata consumed by a
protected collective is authorized to define membership, contribution,
role, or validation-to-use state.

\subsection{Mitigation Results}
\begin{table}[t]
  \centering
  \caption{\shortsys{} mitigation summary.}
  \label{tab:cic-final-summary}
  \footnotesize
  \setlength{\tabcolsep}{3.5pt}
  \begin{tabular}{@{}p{0.60\columnwidth}p{0.32\columnwidth}@{}}
    \toprule
    Outcome & Result \\
    \midrule
    \multicolumn{2}{@{}l}{\textit{Reference preservation}} \\
    Core attack observations mismatch trusted reference & 102/102 \\
    Core hardened executions preserve it & 102/102 \\
    Secondary sync instances preserved & 13/13 \\
    \addlinespace[1pt]
    \multicolumn{2}{@{}l}{\textit{Hardened outcome accounting}} \\
    Explicit mismatch detections & 83 \\
    Prevention-only outcomes & 32 \\
    Core plus secondary hardened outcomes & 115 \\
    \addlinespace[1pt]
    \multicolumn{2}{@{}l}{\textit{Cost and footprint}} \\
    Runtime comparisons & 51 \\
    Cheap-contract timing delta & median $-0.20\%$; range $-3.77$--$4.00\%$ \\
    Cheap-contract p95 delta & median $-0.16\%$; max $11.14\%$ \\
    Resource-profile targets & 17 \\
    Median / p90 / max registers & 24 / 28 / 34 \\
    Nsight retrieval registers / occupancy & 29 vs. 29; 16.26\% vs. 16.26\% \\
    Stack or spill targets & 0 \\
    \bottomrule
  \end{tabular}
\end{table}

\shortsys{} preserves the trusted reference for 102/102 core defined-behavior
instances. It also preserves the 13 secondary synchronization instances under
hardening. Eighty-three hardened executions raise explicit detections; the
remaining hardened cases prevent the failure by recomputing trusted state or
consuming a trusted contract. The 83 detection outcomes and 32 prevention-only
outcomes partition the 115 hardened core-plus-secondary instances: 102
CUDA-defined core instances plus 13 synchronization-sensitive instances.

Simple hardening baselines cover pieces of the problem. Range checks reject
malformed lane ids, widths, and counts, but miss range-valid values with invalid
authority. Descriptor freezing closes one temporal window. Predicate
recomputation protects the contribution path. Full-call-mask predicate gating
keeps examples CUDA-defined. \shortsys{} composes these contract-specific
building blocks into a full binding relation over the named fields. In the
ablation, each partial defense blocks the field it binds; full \shortsys{}
preserves all 133 overlapping contract-family observations
(Table~\ref{tab:cic-ablation}).

\subsection{Cost and Limits}
\shortsys{} checks a semantic relation that range checking cannot express. A
source lane can be between 0 and 31 while violating the role policy. A mask can
contain only legal bits while naming the wrong authority set. A descriptor can
be range-checked while stale at commit time.

\shortsys{} cost follows the trusted-source cost. Cheap public-shape
predicates, canonical roles, frozen descriptors, and local epoch checks are
lightweight in the compact harnesses. Predicate recomputation is the expensive
case because the trusted predicate must be reconstructed at authorization time.

Across 51 compact benchmark overhead comparisons, cheap-contract \shortsys{}
variants have median deltas inside the compact timing interval summarized in
Appendix~\ref{app:mitigation-footprint}. These CUDA-event timings characterize
the compact harnesses; Appendix~\ref{app:mitigation-footprint} also lists
warmups, repeats, baselines, static resource checks, and Nsight smoke-profile
details. In the direct-trigger slice, the median runtime delta is $-0.20\%$,
the mean runtime delta is $0.03\%$, and the median p95 delta is $-0.16\%$; the
largest p95 delta in that slice is $11.14\%$. The static resource profile
covers 17 targets and 52 kernel entries, with median/p90/max registers of
24/28/34 and no stack or spill targets. The Nsight retrieval smoke profile
reports equal register count and achieved occupancy for benign and hardened
variants in the compact harness; DRAM throughput utilization is 1.20\% versus
1.27\%, and L2 hit rate is 67.90\% versus 67.75\%. These profiler numbers are a
resource-smoke check for one compact guard, not a production throughput claim.

The expensive recompute guard is near the compact timing interval at 0 and 64
validation rounds, and rises to about 45\% and 228\% over the no-defense attack
path at 256 and 1024 rounds. We interpret \shortsys{} as a low-overhead
discipline for cheap or amortized contracts, with explicit cost when the
trusted contract is expensive to reconstruct.

%% file: sections/09_discussion_limitations.tex
\section{Discussion and Limitations}
\label{sec:discussion}

\subsubsection*{Scope}
CSC applies when reachable metadata defines membership, contribution, role, or
temporal binding for a security-relevant CUDA collective and changes the
protected observation relative to the trusted reference. Nearby scalar or
descriptor bugs are out of scope unless they become authority-bearing metadata
at such a collective site.

\subsubsection*{Complementary defenses}
Memory safety prevents some causes of participation corruption, especially
out-of-bounds writes into masks or predicates. It leaves open whether in-bounds,
range-valid, host-controlled, stale, or adversarial metadata is the authority a
collective should represent. Control-flow defenses protect
execution targets such as returns, indirect branches, and code pointers.
Data-flow integrity or taint tracking could help if the policy names the
relevant objects. Our contribution is to identify those CUDA collective objects
and their use sites.

\subsubsection*{CUDA-definedness and architecture scope}
Some CUDA synchronization mistakes enter undefined-behavior territory. We
separate those cases from the main result: the evaluation uses 102 CUDA-defined
core instances, reports 13 bounded synchronization-sensitive instances
separately, and uses no UB-sensitive instances for the core claim. The
experiments and wrappers are limited to NVIDIA CUDA. Additional NVIDIA replays
provide platform-robustness evidence within that ecosystem, not multi-vendor
coverage.

\subsubsection*{Scale, checker, and cost}
The workloads are compact and deterministic so that participation failures and
trusted references remain auditable. Full framework integration, larger
application benchmarks, and production-throughput studies require separate
evaluation. The review-surface checker triages candidate sites for manual
review, not vulnerability prevalence. \shortsys{} cost depends on how the
trusted contract is obtained: cheap or amortized contracts are lightweight in
our compact harnesses, while expensive predicate recomputation can be costly.

%% file: sections/10_related_work.tex
\section{Related Work}
\label{sec:related}

CSC identifies participation metadata as the security object for CUDA
collectives. It situates this object between non-control-data/DFI work on
protected scalar state, GPU memory-safety work on corruptible device state,
SIMT verification on masks, barriers, and divergence, and CUDA library idioms
for membership, contribution, role, and validation-to-use binding.

\subsubsection*{Non-control-data attacks and DFI}
Non-control-data attacks and data-oriented programming show that security can
fail even when control flow is preserved~\cite{chen2005noncontrol,hu2016dop}.
Collective Semantic Corruption is a CUDA collective-decision instance of that
lesson: the corrupted object is participation metadata that determines which
lanes a decision represents, what they contribute, and which roles they occupy.
Control-flow and code-pointer integrity protect a complementary control
object~\cite{abadi2005cfi,kuznetsov2014cpi}; CSC assumes that object remains
intact. DFI, taint, and modular non-control-data defenses can help when a
policy names the protected object
~\cite{castro2006dfi,schlesinger2014modular}. CSC identifies the CUDA
collective objects such a policy would name: membership, contribution, role,
and temporal-binding state at collective use sites.

\subsubsection*{GPU exploitation and memory safety}
GPU exploitation work shows that CUDA memory bugs can lead to code injection,
ROP, and code reuse~\cite{guo2024gpu,roels2025cuda}, connecting GPUs to the
broader memory-corruption and code-reuse literature
~\cite{shacham2007rop,bletsch2011jop,szekeres2013sok}. GPU memory-safety
systems build on a broader line of spatial and temporal memory-safety tools
and compiler defenses
~\cite{serebryany2012asan,akritidis2009baggy,nagarakatte2009softbound,
nagarakatte2010cets}. GPUShield and CUSAFE address CUDA corruption causes
through bounds checking or memory-corruption detection
~\cite{lee2022gpushield,lu2026cusafe}. These defenses can prevent some
corruptions that feed our attack instances. They leave open whether
range-valid or in-bounds metadata is the right authority for a
security-relevant collective.

\subsubsection*{GPU data manipulation and side channels}
GPU-resident data manipulation and GPU side-channel work show additional
security consequences beyond control-flow compromise
~\cite{naghibijouybari2018rendered,wang2024gpuzip,zhao2024owl}. CSC focuses on
the collective instruction reached by the kernel, where program-visible
non-control data changes participation authority.

\subsubsection*{SIMT semantics and verification}
Formal and verification work on SIMT execution, races, and barrier divergence
treats masks, predication, and synchronization as correctness-critical
~\cite{habermaier2012simt,betts2012gpuverify,li2012gklee,
li2014symbolicrace,zheng2014gmrace,eizenberg2017barracuda,price2015oclgrind,
wu2019cudaBugs,wu2019aucs,wu2020simulee,joshi2021gpurepair}. We reinterpret a
subset of those objects under an adversarial security question: can mutable
participation metadata make a CUDA-defined collective represent invalid
authority?

\subsubsection*{CUDA guidance and library idioms}
NVIDIA documentation describes warp-level primitives, masks, participation
requirements, and Cooperative Groups
~\cite{nvidiaCudaGuide,nvidiaPtxIsa,nvidiaCooperativeGroups,
linCudaWarpPrimitives,nvidiaBestPractices}. Prefix sums, scans, reductions,
and compaction have a long data-parallel history
~\cite{blelloch1990prefix,sengupta2007scan,sengupta2008scan}. CUDA libraries
such as CUB, Thrust, CUDPP, ModernGPU, CUTLASS, cuDNN, Gunrock, and FAISS use
these idioms in tiled kernels, graph frontiers, filtering, and top-\(k\)
selection~\cite{nvidiaCub,nvidiaThrust,cudpp,moderngpu,merrill2016scan,
kerr2017cutlass,chetlur2014cudnn,wang2016gunrock,johnson2021faiss}.
Heterogeneous benchmark suites such as Rodinia provide additional workload
context~\cite{che2009rodinia}. These systems provide idiom motivation without
serving as vulnerability reports.

%% file: sections/11_conclusion.tex
\section{Conclusion}
\label{sec:conclusion}

CUDA collective decisions expose participation metadata as security-relevant
non-control data. That metadata determines which lanes are represented, what
they contribute, which lane may speak, and whether validated state remains
bound at commit. We characterized \emph{Collective Semantic Corruption} as a
non-control-data attack family in which CUDA-conforming collectives reach the
intended site and compute over invalid participation authority. Our NVIDIA CUDA
evidence is a contract-conformance result across primitive, trigger-channel,
compact workload, reduced-idiom, and admission-guard harnesses, with
CUDA-defined and synchronization-sensitive instances separated.
\sys{} gives the corresponding binding discipline: derive, check, recompute,
or freeze participation metadata before a security-relevant collective consumes
it. For CUDA collective decisions, non-control data includes the values being
computed and who is allowed to count.

%% file: sections/appendix.tex
\setlength{\textfloatsep}{6pt plus 1pt minus 2pt}
\setlength{\floatsep}{5pt plus 1pt minus 1pt}
\setlength{\intextsep}{5pt plus 1pt minus 1pt}
\setlength{\dbltextfloatsep}{6pt plus 1pt minus 2pt}
\setlength{\dblfloatsep}{5pt plus 1pt minus 1pt}
\setlength{\abovecaptionskip}{3pt}
\setlength{\belowcaptionskip}{0pt}
\setlength{\emergencystretch}{1.5em}
\footnotesize

\section{Reproducibility and Scope}
\label{app:reproducibility-pointers}

An anonymized Docker-first artifact is available at
\url{https://anonymous.4open.science/r/warp-corruption/}. It contains CUDA
sources, model and CUDA fuzzers, replay wrappers, validators, ledgers, and
frozen paper-result summaries. The artifact is organized around one container
and five make targets: \texttt{validate}, \texttt{cpu-smoke},
\texttt{cuda-smoke}, \texttt{performance-smoke}, and
\texttt{full-cuda-replay}. CPU validation and package checks do not require
CUDA; CUDA smoke, full replay, and performance smoke require an NVIDIA CUDA
host with the NVIDIA Container Toolkit.

\begin{table}[!htbp]
  \centering
  \scriptsize
  \caption{Artifact map. These files define how to reproduce and audit the
  paper-facing results. Ledger filenames live under
  \texttt{benchmarks/collective-integrity/} unless noted.}
  \label{tab:artifact-map}
  \setlength{\tabcolsep}{3pt}
  \resizebox{\columnwidth}{!}{%
  \begin{tabular}{@{}p{0.47\columnwidth}p{0.48\columnwidth}@{}}
    \toprule
    Artifact path & Role \\
    \midrule
    \texttt{ARTIFACT\_MANIFEST.json} & top-level audit map \\
    \texttt{RUNBOOK.md}, \texttt{Makefile} & command sequence and make targets \\
    \texttt{EXPECTED\_RESULTS.md} & expected replay and validation outcomes \\
    \texttt{core\_instances.csv} & 102 CUDA-defined core instances \\
    \texttt{per\_case\_ledger.csv} & case-family provenance ledger \\
    \texttt{evaluation\_accounting.json} & run and denominator accounting \\
    \texttt{paper-results/summaries/} & frozen paper-facing summaries \\
    \texttt{results/} & host-local scratch output for fresh runs \\
    \bottomrule
  \end{tabular}
  }
\end{table}

The intended scope remains program-visible participation metadata, compact
benchmark and application-guard harnesses, modeled checker findings, and
NVIDIA CUDA replay data. The paper-facing same-site replay records 59/59
paired direct attacks with matched benign twins and 59/59 SASS/PTX
binary-context availability. Architecture replay is NVIDIA platform robustness
evidence, not a multi-vendor or production-throughput claim: the local RTX
4070 Laptop/sm\_89 run is reported separately, while V100/sm\_70, T4/sm\_75,
A100/sm\_80, and H100/sm\_90 summaries cover the active core replay on four
additional NVIDIA targets. The artifact intentionally excludes the paper
source, editorial worklogs, non-public report drafts, local checkout caches,
and machine-specific remote-host configuration.

\section{Ethical Considerations}
\label{app:ethical-considerations}

This work studies security-relevant misuse of CUDA collective participation
metadata. The experiments use local CUDA harnesses, reduced idiom bridges,
public source code, and reproducible artifact scripts; they do not involve
human subjects, personal data, credentials, private datasets, or live
third-party services. Checker results are treated as review-surface evidence,
not vulnerability prevalence or public bug claims.

The work has limited dual-use risk because it characterizes attack patterns
and provides compact reproducer code. We mitigate that risk by framing the
results as contract violations, including corresponding hardened variants,
and omitting private reports, credentials, host-specific configuration, and
local third-party corpora from the public artifact. Where follow-up identified
a third-party public-API output-integrity issue, we privately disclosed it to
the maintainers before submission and avoid publishing exploit-enabling
operational details beyond the minimal scientific description needed to
support the paper's claims.

\section{Run Accounting}
\label{app:run-accounting}

This appendix is a guide to the artifact ledgers, not a replacement for them.
The main paper uses the counts in Table~\ref{tab:evaluation-claims}; the
tables below map those counts to logged executions, counted instances, support
experiments, and scope boundaries.

\begin{table}[!htbp]
  \centering
  \scriptsize
  \caption{Run-accounting audit trail. Records are logged executions; instances
  and cases are claim-specific subsets.}
  \label{tab:appendix-evaluation-accounting}
  \setlength{\tabcolsep}{3pt}
  \resizebox{\columnwidth}{!}{%
  \begin{tabular}{@{}p{0.18\columnwidth}p{0.38\columnwidth}p{0.10\columnwidth}p{0.25\columnwidth}@{}}
    \toprule
    Quantity & Meaning & Count & Use \\
    \midrule
    Full run records & logged benign, attack, hardened, and control runs & 1251 & traceability \\
    Architecture-replay records & replay subset used for platform-robustness checks & 340 & robustness \\
    CIC/eval records & \shortsys{} mitigation, benchmark, and checker records & 367 & mitigation \\
    Benchmark records & benchmark-only records in the mitigation set & 345 & case coverage \\
    Benchmark cases & distinct kernel scenarios in the main benchmark set & 25 & coverage summary \\
    Core attack instances & CUDA-defined non-control-data attack instances & 102 & core result \\
    Secondary sync instances & bounded synchronization-divergence instances reported separately & 13 & secondary evidence \\
    UB-sensitive instances & excluded from core because they require UB-sensitive behavior & 0 & boundary \\
    Negative controls & explicit control records plus retrieval non-triggering controls & 12 + 144 & specificity \\
    \bottomrule
  \end{tabular}
  }
\end{table}
\begin{table}[!htbp]
  \centering
  \scriptsize
  \caption{Grouped breakdown of the 102 CUDA-defined core instances.}
  \label{tab:core-instance-breakdown}
  \setlength{\tabcolsep}{2pt}
  \resizebox{\columnwidth}{!}{%
  \begin{tabular}{@{}lrrrr@{}}
    \toprule
    Authority view & Primitive families & Trigger channels & Core instances & Hardened preserved \\
    \midrule
    Membership & 4 & 2 & 26 & 26/26 \\
    Contribution & 6 & 2 & 34 & 34/34 \\
    Role & 2 & 2 & 12 & 12/12 \\
    Temporal & 2 & 2 & 12 & 12/12 \\
    Mixed membership/contribution & 5 & 2 & 18 & 18/18 \\
    \midrule
    Total & 16 & 2 & 102 & 102/102 \\
    \bottomrule
  \end{tabular}
  }
  \vspace{2pt}
  \parbox{\columnwidth}{\scriptsize Primitive-family counts overlap across
  authority views; the Total row reports deduplicated primitive families. Core
  instances are disjoint and sum to 102. Trigger-channel counts report labels
  represented in the grouped core ledger; Section~\ref{sec:evaluation} reports
  the three threat-model channels separately.}
\end{table}

\begin{table}[!htbp]
  \centering
  \scriptsize
  \caption{Claim audit map. CSV ledgers contain full per-case provenance.}
  \label{tab:appendix-claim-audit}
  \setlength{\tabcolsep}{3pt}
  \resizebox{\columnwidth}{!}{%
  \begin{tabular}{@{}p{0.24\columnwidth}p{0.32\columnwidth}p{0.36\columnwidth}@{}}
    \toprule
    Claim & Audit source & Boundary \\
    \midrule
    Core defined-behavior result & \texttt{core\_instances.csv}; Tables~\ref{tab:appendix-evaluation-accounting} and~\ref{tab:core-instance-breakdown} & conforming call masks; no UB-sensitive core instances \\
    Trigger separation & channel-tagged benchmark records; overflow-trigger overlay & causes differ, consequence is the same authority violation \\
    Application impact & retrieval service table in main text; compact guard records & compact harnesses, not production vulnerability prevalence \\
    Tool and site boundary & sanitizer checks, paired same-site runs, marker traces & boundary evidence, not complete trace equivalence \\
    \shortsys{} mitigation and cost & Tables~\ref{tab:cic-ablation} and~\ref{tab:timing-methodology} & named-field binding, not policy inference or memory safety \\
    NVIDIA scope & Appendix~\ref{app:reproducibility-pointers} & NVIDIA CUDA robustness, not multi-vendor coverage \\
    \bottomrule
  \end{tabular}
  }
\end{table}

\section{Evaluation Support Tables}
\label{app:evaluation-support-tables}

These tables expand the evaluation without adding vulnerability claims. They
show which trusted source is compared with which mutable metadata and record
the exact fuzzer counts used by Section~\ref{sec:evaluation}. Reduced idiom
bridges, negative controls, and checker audit questions are summarized in the
main text and recorded in the artifact.

\begin{table}[!htbp]
  \centering
  \scriptsize
  \caption{Trusted contract source versus untrusted participation metadata.}
  \label{tab:trusted-vs-untrusted}
  \setlength{\tabcolsep}{2pt}
  \begin{tabular}{@{}p{0.16\columnwidth}p{0.22\columnwidth}p{0.20\columnwidth}p{0.22\columnwidth}p{0.13\columnwidth}@{}}
    \toprule
    Case & Trusted source & Untrusted field & Why it looks safe & Violation \\
    \midrule
    Tail validation & public \texttt{lane<n} & descriptor mask & descriptor is range checked & lane dropped \\
    Sparse gather & verified row length & row mask / offset & sparse descriptors are normal inputs & wrong members \\
    Shuffle leader & first valid lane & leader lane & source lane is range valid & wrong speaker \\
    Retrieval guard & recomputed validity & stored valid flags & previous stage wrote them & invalid admit \\
    TOCTOU & checked copy & reloaded descriptor & same pointer was checked earlier & check/use drift \\
    \bottomrule
  \end{tabular}
\end{table}

\begin{table}[!htbp]
  \centering
  \scriptsize
  \caption{Fuzzer methodology and exact per-dimension counts.}
  \label{tab:fuzzing-methodology}
  \setlength{\tabcolsep}{3pt}
  \resizebox{\columnwidth}{!}{%
  \begin{tabular}{@{}llrrrrrrr@{}}
    \toprule
    Fuzzer & Dimension & Generated & Valid & Attack-divergent & Control states & \shortsys{} preserved divergent states & CIC FN & Control FP \\
    \midrule
    Model, seed 424242 & Membership & 16,425 & 16,425 & 13,098 & 3,327 & 13,098 & 0 & 0 \\
    Model, seed 424242 & Contribution & 16,345 & 16,345 & 13,033 & 3,312 & 13,033 & 0 & 0 \\
    Model, seed 424242 & Role & 16,244 & 16,244 & 12,980 & 3,264 & 12,980 & 0 & 0 \\
    Model, seed 424242 & Temporal & 16,522 & 16,522 & 13,229 & 3,293 & 13,229 & 0 & 0 \\
    Model total & All dimensions & 65,536 & 65,536 & 52,340 & 13,196 & 52,340 & 0 & 0 \\
    \midrule
    CUDA, seed 1337 & Membership & 16,384 & 16,384 & 13,129 & 3,255 & 13,129 & 0 & 0 \\
    CUDA, seed 1337 & Contribution & 16,384 & 16,384 & 13,164 & 3,220 & 13,164 & 0 & 0 \\
    CUDA, seed 1337 & Role & 16,384 & 16,384 & 13,130 & 3,254 & 13,130 & 0 & 0 \\
    CUDA, seed 1337 & Temporal & 16,384 & 16,384 & 13,199 & 3,185 & 13,199 & 0 & 0 \\
    CUDA total & All dimensions & 65,536 & 65,536 & 52,622 & 12,914 & 52,622 & 0 & 0 \\
    \bottomrule
  \end{tabular}
  }
\end{table}

\section{Same-Site Preservation Protocol}
\label{app:same-site-protocol}

The same-site checks compare paired benign and attack executions for the same
case, trigger class, launch shape, and collective site. The metadata payload
changes; the intended collective site and primitive do not. The current paired
core replay records binary context for every paired direct attack: benign twin
reference pass 59/59, attack twin reference fail 59/59, same case/site/trigger
pairing 59/59, and SASS/PTX hash evidence 59/59. A separate marker harness
covers one kernel per authority dimension. Membership, contribution, role, and
temporal traces preserve the same marker sequence in 4/4 benign/attack pairs
while the attack decision diverges and hardened mode detects in 4/4. These
checks support the same-site boundary; they are not dynamic trace-equivalence
or family-wide trace evidence.

\section{Case-Study Contract Details}
\label{app:case-study-contracts}

The case-study tables spell out the trusted-source/corrupted-object relation
for the compact application guards and reduced idiom bridges. They are contract
audits, not reports of vulnerabilities in upstream libraries.

\begin{table}[!htbp]
  \centering
  \scriptsize
  \caption{Contract view for representative application-level and reduced
  idiom bridges.}
  \label{tab:case-study-audit}
  \setlength{\tabcolsep}{3pt}
  \resizebox{\columnwidth}{!}{%
  \begin{tabular}{@{}p{0.20\columnwidth}p{0.24\columnwidth}p{0.22\columnwidth}p{0.24\columnwidth}@{}}
    \toprule
    Case study & Trusted contract source & Corrupted object & No-defense $\rightarrow$ \shortsys{} result \\
    \midrule
    Retrieval admission guard & recomputed candidate validity & prior-stage stored validity mask & invalid candidate admitted $\rightarrow$ rejected/detected \\
    Queue-rank frontier & canonical active-mask rank and frontier window & lane rank / output slot metadata & invalid frontier admitted $\rightarrow$ rejected/detected \\
    Multi-kernel temporal guard & descriptor count and epoch frozen after validation & reloaded count / epoch & stale descriptor accepted $\rightarrow$ drift detected \\
    Warp/reduction idioms & recomputed validity, group membership, or expected subwarp domain & include bit, reduce mask, width, source lane & poisoned aggregate or wrong representative $\rightarrow$ reference restored \\
    Batchnorm / packet / frontier guards & finite/range predicate, full segment membership, public frontier size & stored validity, segment mask, frontier count & invalid sample, packet, or frontier admitted $\rightarrow$ rejected/detected \\
    \bottomrule
  \end{tabular}
  }
\end{table}

\section{\shortsys{} Support Tables}
\label{app:cic-support-tables}

These tables document the \shortsys{} wrapper surface and ablation views.
Field-view observations overlap and should not be summed with the core
instance set.

\begin{table}[!htbp]
  \centering
  \scriptsize
  \caption{\shortsys{} API surface and field coverage.}
  \label{tab:cic-contracts}
  \setlength{\tabcolsep}{3pt}
  \resizebox{\columnwidth}{!}{%
  \begin{tabular}{@{}p{0.17\columnwidth}p{0.16\columnwidth}p{0.31\columnwidth}p{0.22\columnwidth}p{0.13\columnwidth}@{}}
    \toprule
    Helper / contract & Field & Enforcement rule & Stops & Limit \\
    \midrule
    expect/check mask & membership & compute or compare expected participants & mask substitution, group confusion & wrong trusted source \\
    all/reduce & contribution & use trusted predicates or recomputed contributions & vote spoofing, poisoned reductions & expensive recompute \\
    shuffle binding & role & bind source, width, leader, or rank & wrong speaker or output owner & ambiguous policy \\
    commit binding & temporal & commit with validated descriptor, epoch, or copy & TOCTOU and stale reload & pre-contract corruption \\
    sync binding & reachability & check intended barrier reachability & barrier participation drift & sync policy required \\
    \bottomrule
  \end{tabular}
  }
\end{table}

\begin{table}[!htbp]
  \centering
  \scriptsize
  \caption{\shortsys{} contract-family ablation evidence. Field-view
  observations overlap; the final line gives the deduplicated count used in the
  text.}
  \label{tab:cic-ablation}
  \setlength{\tabcolsep}{2pt}
  \resizebox{\columnwidth}{!}{%
  \begin{tabular}{@{}lrrrr@{}}
    \toprule
    Field view & Unique cases & Field-view obs. & Attack violations & Hardened preserved \\
    \midrule
    Membership & 8 & 34 & 34/34 & 34/34 \\
    Mask provenance & 7 & 31 & 31/31 & 31/31 \\
    Predicate revalidation & 9 & 40 & 40/40 & 40/40 \\
    Temporal stability & 3 & 14 & 14/14 & 14/14 \\
    Barrier reachability & 1 & 7 & 7/7 & 7/7 \\
    Role/domain authority & 5 & 14 & 14/14 & 14/14 \\
    \midrule
    Deduplicated observations & -- & 133 & 133/133 & 133/133 \\
    \bottomrule
  \end{tabular}
  }
\end{table}

\section{Mitigation Footprint Details}
\label{app:mitigation-footprint}

A reduced workload/idiom footprint study covers ten source files used by the
reduced-idiom bridges and compact application-level guards. Those files contain
18 \shortsys{} helper calls across 3769 LOC, with a median of two helper calls
per file. This is a reduced-idiom deployability estimate, not an upstream
patch.

We compile 17 representative targets with \texttt{ptxas -v}, covering 52
kernel entry functions when the Gunrock checkout is present. The maximum
reported use is 34 registers per entry, with median/p90 per-target maxima of
24/28 registers and with 0 stack bytes and 0 spill load/store bytes. The
frontier guard uses 2048 bytes of static shared memory; the multi-kernel
temporal guard uses 28 registers; the queue-rank bridge and API-path
\texttt{Queue::next()} validation each use 24 registers. This is a static
resource check, not an occupancy study.

An optional Nsight Compute smoke profile on benign and hardened
retrieval-guard launches reports equal 29-register/thread usage, about 16.3\%
achieved occupancy, 1.20/1.27\% DRAM-throughput utilization, and 67.90/67.75\%
L2 hit rate in the compact one-block harness.

Timing records use CUDA events around compact harness launches after warmups.
The emitted benchmark JSON records warmups, repeats, baseline path, median
delta, and expensive-recompute setting when available. Clocks are not pinned,
so cheap-contract deltas inside the reported run-to-run interval are
smoke-level timing evidence rather than production-throughput claims. The
expensive recompute guard is the explicit boundary case: it stays near the
compact timing interval at 0 and 64 validation rounds, then rises to 45.08\%
and 228.49\% over the no-defense attack path at 256 and 1024 rounds. Public
shape predicates, canonical roles, frozen descriptors, and epoch checks are
low-cost sources; predicate recomputation and debug logging are the
variable-cost cases.

\begin{table}[!htbp]
  \centering
  \scriptsize
  \caption{Timing methodology and compact cost summary. Negative deltas are
  treated as run-to-run timing variation, not speedups.}
  \label{tab:timing-methodology}
  \setlength{\tabcolsep}{3pt}
  \resizebox{\columnwidth}{!}{%
  \begin{tabular}{@{}p{0.20\columnwidth}rrrrp{0.18\columnwidth}p{0.17\columnwidth}p{0.18\columnwidth}@{}}
    \toprule
    Benchmark group & Process warmups & Samples & Device warmups & Device repeats & Baseline & Median / range & Notes \\
    \midrule
    Direct cheap-contract slice & 3 & 15 & 100 & 2000 & benign vs. hardened & median $-0.20\%$; range $-3.77$--$4.00\%$ & 17 compact comparisons; CUDA-event timing \\
    New direct idiom/app comparisons & 3 & 15 & 100 & 2000 & benign vs. hardened & $-0.87$--$1.88\%$ & FAISS-style, ArrayFire-style, retrieval pipeline guards \\
    Pipeline scale direct & 3 & 15 & 100 & 2000 & benign vs. hardened & $-0.76$--$0.63\%$ & 8, 16, and 32 candidates \\
    Expensive recompute, 0 rounds & n/a & 1 summary & 200 & 2000 & no-defense attack path & $-0.80\%$ & near compact timing interval \\
    Expensive recompute, 64 rounds & n/a & 1 summary & 200 & 2000 & no-defense attack path & $0.03\%$ & near compact timing interval \\
    Expensive recompute, 256 rounds & n/a & 1 summary & 200 & 2000 & no-defense attack path & $45.08\%$ & costly-contract boundary \\
    Expensive recompute, 1024 rounds & n/a & 1 summary & 200 & 2000 & no-defense attack path & $228.49\%$ & costly-contract boundary \\
    \bottomrule
  \end{tabular}
  }
\end{table}